\documentclass{emulateapj}
\usepackage{amsmath,amssymb}
\usepackage{epstopdf}
\shorttitle{SEP Acceleration by a Shock Accompanying CME}
\shortauthors{Petukhova et al.}

\begin{document}

\title{Solar Energetic Particle Acceleration by a Shock Wave Accompanying 
a Coronal Mass Ejection in the Solar Atmosphere}

\author{A. S. Petukhova, I. S. Petukhov, S. I. Petukhov, L. T. Ksenofontov}
\affil{Yu. G. Shafer Institute of Cosmophysical Research and Aeronomy, SB RAS
\\ 677980, 31 Lenin Ave., Yakutsk, Russia}

\email{i\_van@ikfia.sbras.ru}

\begin{abstract}
Solar energetic particles acceleration by a shock wave accompanying a coronal 
mass ejection (CME) is studied. The description of the accelerated particle 
spectrum evolution is based on the numerical calculation of the diffusive 
transport equation with a set of realistic parameters. The relation between the 
CME and the shock speeds, which depend on the initial CME radius, is 
determined. Depending on the initial CME radius, its speed, and the magnetic 
energy of the scattering Alfv\'{e}n waves, the accelerated particle spectrum is 
established during 10--60 minutes from the beginning of CME motion. The 
maximum energies of particles reach 0.1~--~10 GeV. The CME radii of 3~--~5 
$R_\odot$ and the shock radii of 5~--~10 $R_\odot$ agree with observations. 
The calculated particle spectra agree with the observed ones in events 
registered by ground-based detectors if the turbulence spectrum in the solar 
corona significantly differs from the Kolmogorov one.
\end{abstract}


\section{Introduction}

In the last two decades, it has been found that solar energetic particles (SEPs) 
in gradual events are generated from particle acceleration by shock waves 
accompanying coronal mass ejections (CMEs) \citep[see reviews by][and 
references 
therein]{rea99,lee05}. It was confirmed by both the correspondence of the solar 
corona composition with the SEP composition, on average, and a close 
association of 
gradual events with CMEs. Observations show that SEP generation occurs in the 
vicinity of the Sun. SEP injection into interplanetary space begins about 
0.2--1 hr after the beginning of CME motion \citep{kah94, kru00}.
At the beginning of SEP injection, the CME radius is 3~--~5 ${R_\odot }$, where 
${R_ \odot }$ is 
the solar radius. The injection of SEPs with ${\varepsilon _k} > 50$ MeV 
nucleon$^{-1}$ 
has a characteristic behavior: a fast increase to the maximum value is followed 
by an exponential decrease. Here, ${\varepsilon _k}$ is the kinetic energy per 
nucleon. At the same time, the higher ${\varepsilon _k}$ is, the earlier the
injection reaches its maximum value, and the characteristic decreasing timescale 
${t_*}\propto 1/{\varepsilon _k}$. In the events registered by ground-based 
detectors (GLE; ${\varepsilon _k} > 1$ GeV) the injection reaches its maximum 
value when the CME radius is within 4~--~30 ${R_ \odot }$ and ${t_*}\sim 1-10$ 
hr. The SEPs injected during a decreasing phase are generated in the regions 
within 5~--~30 $R_\odot$. All observations of SEPs in gradual events imply that 
particle acceleration occurs in the region from 2 to 30 ${R_ \odot }$ 
\citep{lee05}. The particles effectively accelerates at the shock front because 
of a 
rather high level of background and self-consistently generated turbulence, as 
well as the high CME speed.

In interplanetary space, acceleration efficiency significantly decreases since 
the level of background turbulence \citep{bie94} is much less than in the solar 
corona. Thus, the SEPs accelerated in the solar corona escape from the vicinity 
of the shock front. The temporal behavior of SEP flux in gradual events at the 
Earth's orbit depends on particle energy. The flux of SEPs with energies $(10 < 
{\varepsilon _k} < 50)$ MeV nucleon$^{-1}$ after increasing remains almost 
constant 
until the shock front reaches the Earth's orbit--- the region of a 
plateau in 
the temporary event profile \citep{rea90}. The flux of SEPs with ${\varepsilon 
_k} < 10$ MeV nucleon$^{-1}$ after the plateau region increases, reaching the 
maximum 
value at the shock front --- the region of energetic storm particles (ESPs) 
\citep{lee83}. After passing the shock front, SEP spectra exponentially decrease 
with time at scales ${t_*}\sim 1 - 10$ hr, retaining their power-law shape---
the region of invariant SEP spectrum decrease \citep{rea97}.

There are different approaches in theoretically describing SEP generation 
and propagation. \citet{zank00} presented the model of proton acceleration at 
the 
expanding shock wave; it is an analogue of the model used by \citet{bog83} to 
describe the particle acceleration in supernova remnants. In the solution, the 
accelerated particle spectrum at a plane shock is used, which is subsequently 
modified by diffusion, convection, and adiabatic energy losses in the expanding 
spherical CME volume. They assume that the generation of self-consistent 
turbulence is very effective that the diffusion coefficient reaches the Bohm 
limit. The model results and observational properties of some gradual events are 
generally in agreement. An analytical theory of ion acceleration at spherical 
shock waves was presented by \citet{lee05}. The theory is based on works 
\citep{lee83, gor99} containing coupled ion acceleration and wave generation at 
the plane shock front. The achievement of this new theory is that it includes a 
continuous transition from the region with frequent particles scattering near 
the shock front to the region with almost free propagation far from the shock 
front. The theory's simplification is reached by using of stationary particle 
spectra and self-consistent waves during the entire time of the shock movement. 

Particle acceleration based on the numerical solution of the diffusive 
transport equation under the solar corona conditions was considered in the 
works of \citet{bt03,bt13}. Both linear and nonlinear cases of the problem are 
considered 
with the self-consistent turbulence generation by the accelerated particles. 
The 
parameters of the model are determined from the comparison of the calculated 
particle 
spectra with SEP spectra in gradual events. These models did not take into 
account the following: (1) CME influence on the shock speed, (2) influence of 
the 
region behind the shock front on particle acceleration, and (3) dependence of 
the 
accelerated particle spectra on the initial CME radius in the solar corona.

Here we present the results of a linear model of particle acceleration at an 
expanding spherical shock wave in the the solar corona. The spectra are 
determined by numerically solving the diffusive transport equation. In the
calculations, we use the relation between the CME and the shock speeds 
determined 
by the solution of the gas-dynamic equations. The influence of the region 
behind the 
shock front on particle acceleration is investigated. The dependence of 
particle spectra on the initial CME radius is explored. The model parameters 
are 
determined by comparing the calculated particle spectra with SEP spectra in 
gradual events. An accuracy estimate of the code is given in the Appendix.

\section{Model}

We suggest that the SEP acceleration occurs in the region limited to a few 
solar 
radii \citep{kah94, kru00, rea09}. We also take into account that (1) the 
magnetic field is radial in the acceleration region \citep{sit99}, (2) 
high-energy particles in the solar atmosphere are strongly magnetized, so that 
${\kappa_\perp}/{\kappa_\parallel} \ll 1$, where $\kappa_\perp$ and $\kappa 
_\parallel$ are the diffusion coefficients across and along the magnetic field, 
respectively. In this case, the particle acceleration in the central part of the 
shock weakly depends on its configuration and is the same as in the spherical 
symmetric case. The corresponding transport equation for the isotropic part of 
the distribution function $f (p,r,t)$ is 
\begin{equation}
\frac{{\partial f}}{{\partial t}} = \frac{1}{{{r^2}}}\frac{\partial }{{\partial 
r}}\left( {\kappa {r^2}\frac{{\partial f}}{{\partial r}}} \right) - 
w'\frac{{\partial f}}{{\partial r}} + \frac{1}{{3{r^2}}}\frac{{\partial 
f}}{{\partial r}}\left( {w'{r^2}} \right)p\frac{{\partial f}}{{\partial p}} + Q, 
\label{eq1}
\end{equation}
where $\kappa $ is the diffusion coefficient along the magnetic field; $p$, 
$r$, $t$ are the momentum, radius and time, respectively; $w' = w + {c_c}$ is 
the velocity of the scattering centers; $w$ is the plasma flow velocity; $Q = 
{Q_0}\delta \left( {r - {R_s}} \right)$ is the particle source concentrated at 
the shock front; and ${R_s}$ is the shock front radius. We assume that the 
shock front and the CME (piston) are segments of spherical surfaces with radii 
${R_s},{R_p}$, respectively. Particles are scattered by Alfven waves moving 
along the magnetic field lines in opposite directions. The velocity of the
scattering centers ${c_c}$ in the region ahead of the shock front ($r > {R_s}$) 
is determined as ${c_c} = {c_a}\left( {E_\nu ^ +  - E_\nu ^ - } \right)/{E_\nu 
} = {c_a}{\delta _\nu }$, where $E_\nu ^ + ,E_\nu ^ - $ are the Alfven waves' 
magnetic energy densities over frequency, moving from and to the Sun, 
respectively; $ 
E_\nu =E_\nu^+ +E_\nu^- $; ${c_a} = B/\sqrt {4\pi \rho } $ is the Alfven speed; 
$B$ is the magnetic field strength; $\rho $ is the plasma density; and $\nu $ 
is the frequency. In the region behind the shock front ($r < {R_s}$), the 
directions of 
wave propagation become isotropic ($E_\nu^+  = E_\nu^-$), resulting in ${c_c}(r 
< R_s) = 0$. The plasma velocity $u = {V_s} - w$ and the velocity of the 
scattering 
centers $u' = {V_s} - w'$ in the shock rest frame changes abruptly from the 
values 
of ${u_1} = {V_s} - {w_{|{R_s} + 0}}$ and $u'_1 = {V_s} - w'_{|{R_s} + 0}$ at $r 
= {R_s} + 0$ to ${u_2} = u'_2 = {u_1}/\sigma $ at $r = {R_s} - 0$, where ${V_s}$ 
is the shock speed; $\sigma  = \left( {{\gamma _g} + 1} \right)M_1^2/\left( {2 
+ \left( {{\gamma _g} - 1} \right)M_1^2} \right)$ is the shock compression 
ratio; ${\gamma _g} = 5/3$ is the adiabatic index; ${M_1} = {u_1}/{c_{1s}}$ is 
the 
sonic Mach number; ${c_{1s}} = \sqrt {{\gamma _g}{P_1}/{\rho _1}} $ is the sound 
speed; and ${P_1}$ is the gas pressure. Subscripts 1 and 2 correspond to the 
values 
in the regions ahead of and behind the shock front, respectively. Taking into 
account a step-like change of the parameters, from equation (\ref{eq1}) we can 
obtain the relationship between distribution functions:
\begin{equation}
{\left( {\kappa \frac{{\partial f}}{{\partial r}}} \right)_{|{R_s} + 0}} - 
{\left( {\kappa \frac{{\partial f}}{{\partial r}}} \right)_{|{R_s} - 0}} + 
{Q_0} = \frac{{u'_1 - {u_2}}}{3}p{\left( {\frac{{\partial 
f}}{{\partial p}}} \right)_{|{R_s}}}. \label{eq2}
\end{equation}
The source term ${Q_0} = {u_1}{N_{\rm inj}}\delta \left( {p - {p_{\rm inj}}} 
\right)/(4\pi p_{\rm inj}^2)$ provides an injection of some fraction 
$\eta=N_{\rm inj}/N_{|R_s+0}$ of particles crossing the shock front into the 
acceleration process, where ${N_{|{R_s} + 0}}$ is the upstream particle number 
density; we call $\eta$ the injection rate. For the injection momentum, we use 
$p_{\rm inj} = \lambda m{c_{S2}}$, where $\lambda$ is the numerical factor, $m$ 
is the ion mass, and ${c_{S2}} = {u_1}\sqrt {{\gamma _g}\left( {\sigma  - 1} 
\right) 
+ \sigma /M_1^2} /\sigma $ is the sound speed behind the shock front. 
Kinetic simulations suggest that for parallel shocks, a 
fraction of about $10^{-4} \text{ -- } 10^{-3}$ of the particles crossing the 
shock is injected into the acceleration process with momentum $p_{\mathrm 
inj}=3 \text{ -- } 4 mc_{S2}$ \citep{cap14}. To fit the data by amplitude (see 
Figure 
\ref{f12}), in our calculations we use values of $\eta=10^{-3}$ and $\lambda  = 
3$, which are in good agreement with the results of the kinetic simulations.
The source term determines the amplitude of the distribution function at the 
shock front for a given injection momentum:
\begin{equation}
f\left( {{p_{\rm inj}},{R_s},t} \right) = 3{u_1}\eta {N_{|{R_s} + 0}}/\left( 
{4\pi 
\left( {u'_1 - {u_2}} 
\right)p_{\rm inj}^3} \right). \label{eq3}
\end{equation}
The diffusion coefficient is determined by the relation \citep{lee82}
\begin{equation}
\kappa  = {v^2}{B^2}/\left( {32{\pi ^2}{\omega _B}E\left( {k = \rho _B^{ - 1}} 
\right)} \right), 
\label{eq4}
\end{equation}
where $v$ is the particle speed, ${\omega _B} = eB/(\gamma mc)$ is the 
gyrofrequency, $e$ is the elementary charge, $\gamma $ is the Lorentz factor, 
and $E\left( k \right) = d\left( {\delta {B^2}/8\pi } \right)/(d\ln k) = \nu 
{E_\nu  }$ is the differential density of the Alfven wave magnetic energy over 
the logarithm of the wave number $k$. Particles are scattered by Alfven waves 
whose wave number of is the inverse gyroradius $k = \rho_B^{-1}$.

In our model, the CME and shock front are represented by 
segments of heliocentric spherical surfaces with different radii. However, 
LASCO white-light observations show that the CME shape is not spherical. 
The diffusive shock acceleration is much effective at the quasi-parallel part 
of the shock (the magnetic field is radial), which corresponds to its central 
part. At the periphery of the CME, the shock becomes quasi-perpendicular and 
acceleration is not effective \citep[see, for example,][]{cap14}.
Since particles are strongly magnetized ($\kappa_{\perp} / \kappa_{\parallel} 
\ll 1$), the assumption that the shock front is spherical does not affect 
particle acceleration. Therefore, the heliocentric spherical surface segment of 
the shock in our model corresponds to the region of effective particle 
acceleration.

\section{Solar atmosphere parameters}
For the spatial distribution of the proton number density in the solar 
atmosphere, we use the empirical model by \citet{sit99}: $n\left( 
r \right) = {n_ \odot }{a_1}{e^{{a_2}z}}{z^2}P\left( z \right)$, $P\left( z 
\right) = 1 + {a_3}z + {a_4}{z^2} + {a_5}{z^3}$, where $z = r/{R_ \odot }$, 
${a_1} = 1.3033 \times {10^{ - 3}}$, ${a_2} = 3.6728$, 
${a_3} = 4.8947$, ${a_4} = 7.6123$, ${a_5} = 5.9868$, and ${n_ \odot } = 2.5 
\times {10^8}$ cm$^{-3}$. 

It is assumed that the solar corona density is determined only by protons: $\rho 
= {m_p}n$. To determine the plasma flow velocity, we apply the flow continuity 
condition $w\left( r \right) = {w_ \odot }{n_ \odot }R_ \odot ^2/(n{r^2)}$, 
where $w_\odot=0.52$ km s$^{-1}$, which corresponds to $w\left( {{r_e}} \right) 
= 400$~km~s$^{-1}$, where ${r_e}$ is the astronomical unit. Moreover, we use 
the 
isothermal atmosphere temperature of ${T_ \odot } = 1.6 \times {10^6}$ K and the 
spatial distribution of the radial magnetic field of $B\left( r \right) = {B_ 
\odot }{\left( {{R_ \odot }/r} \right)^2}$, ${B_ \odot 
} = 2.3$ G \citep{hun72}. 

\begin{figure}
\plotone{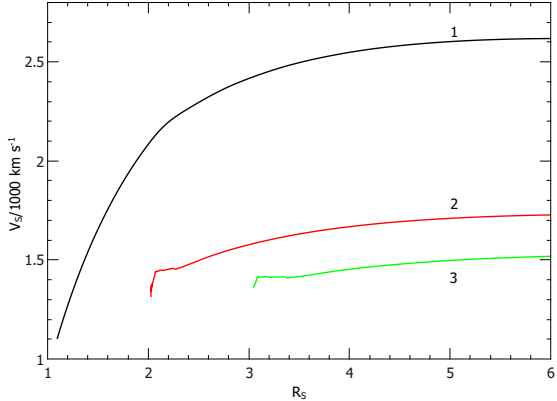} 
\figcaption{Shock speed ${V_s}$ as a function of its radius ${R_s}$ for 
various initial CME radii ${R_{p0}}$. Curves 1, 2, and 3 correspond to 
${R_{p0}}=$ =1, 2,  3, respectively.
The CME speed (${V_p}$) is constant and equals 1000 km s$^{-1}$.
\label{f1}}
\end{figure}

\begin{figure}
\plotone{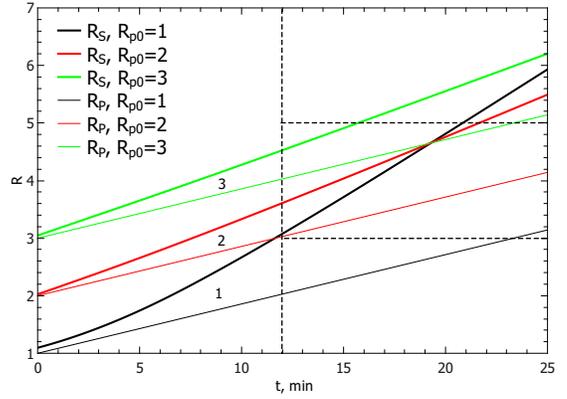} 
\figcaption{Radii of shock fronts (thick lines) and CME (thin lines) as 
functions of time for three initial CME radii. ${V_p} = 1000$ km s$^{-1}$. 
Curves 1, 2, and 3 correspond to ${R_{p0}}=$ =1, 2,  3, respectively.
\label{f2}}
\end{figure}

The CME speed (${V_p}$) is determined from observations. The relation between 
CME and shock speeds is calculated from the linear case of the spherical 
symmetrical problem formulated in \citep{byk96}. 
The CME is represented as that moves with constant velocity from the initial 
time. During the piston movement in the medium, with a given density and plasma 
velocity, the shock radius and 
velocity are determined by calculating the gas-dynamic equations. 
In Figures \ref{f1} and \ref{f2} the calculation results for ${V_p} = 
1000$ km s$^{-1}$ are shown. As we can see, 
the shock speed depends on the initial piston radius. At the beginning, the 
shock speed rapidly increases and then remains almost constant. The 
ratio between the stable speeds can be presented as ${V_s} = {\alpha _s}{V_p}$. 
For ${R_{p0}} =$ 1, 2, 3, the corresponding values are $\alpha _s =$ 
2.6, 1.7, 1.5. Hereinafter, the values of ${R_{p0}}$, ${R_p}$, and ${R_s}$ are 
given in ${R_ \odot }$. Figure \ref{f2} shows the temporary dependence of 
${R_p}$ and ${R_s}$ for three initial CME radii. The flow velocity between 
${R_p}$ and ${R_s}$ for all three cases is close to linear with the radius.

The energy spectrum over frequency in the range $\nu_1 \leq \nu  
\leq \nu_0$ may be expressed as \citep{suz06}
\begin{equation}
{E_\nu } = {E_{v0}}{\left( {\nu /{\nu _0}} \right)^{ - \beta }}{\left( {r/{R_ 
\odot }} \right)^{ - \delta }}, \label{eq5}
\end{equation}
where $\beta  = 1$, ${\nu _1} = 10^{-3}$ Hz, and $\nu_0=5\times 10^{-2}$ Hz. It 
is 
expected that for $\nu  > {\nu _0}$ the spectrum will be softer. In the
calculations, we use Kolmogorov's power-law index $\beta=5/3$, the 
same as in the inertial part of the spectrum in interplanetary space 
\citep{tum95}. The ratio $\nu = k(w+\delta_\nu
c_a)/(2\pi)$ and the condition of particle resonance scattering $k = 
\rho_B^{-1}$ results in $\nu /{\nu _0} = \left( {{p_0}/p} \right) 
(B/B_\odot) (w + {\delta _w}{c_a})/{c_{a \odot }}$, where ${p_0}/{m_p}c = e{B_ 
\odot }{c_{a \odot }}/(2\pi {\nu _0}{m_p}{c^2})$, and ${\delta _\nu } = 
(E_\nu^+ - 
E_\nu^-)/E_\nu$. We determine the diffusion coefficient from the spatial 
energy density of waves with $\nu  \geq \nu_0$, which provides the particle 
scatterings: ${E_w}( \nu  \geq \nu _0 ) = E_{w \odot}/[ 1+( \beta - 1) \ln ( 
\nu 
_0/\nu _1) ]$ and ${\nu _0}{E_{\nu 0}} = (\beta - 1) E_w ( \nu  \geq \nu _0 )$. 
Here $E_{w \odot } = E_w ( \nu  \geq \nu_1)$ is the total energy density, 
which can be found from the ratio ${E_{w \odot }} = {F_ \odot }/{c_{a \odot }}$, 
where ${F_ \odot }$ is the energy flow density of waves moving from the Sun, 
and 
${w_ \odot } \ll {c_{a \odot }}$ is used. In the calculation, we use ${F_ \odot 
} \simeq 5 \times {10^5}$ erg cm$^{-2}$s$^{-1}$) \citep[][and references 
therein]{bt13}. 
Hence, we get the expression for the particle diffusion coefficient 
\begin{equation}
\kappa  = {\kappa _p}{\kappa _r},  \label{eq6}
\end{equation}
where 
\[{\kappa _p} = {\kappa _0}\frac{{{{\left( {p/{m_p}c} \right)}^{3 - \beta 
}}}}{{\sqrt {1 + {{\left( {p/({m_p}c)} \right)}^2}} }},\] 
\[{\kappa _r} = {\left( {r/{R_ \odot }} \right)^{\delta  - 4\beta  
+ 2}}{\left( {\frac{{{w_ \odot }/{c_{a \odot }} + {\delta _\nu 
}n_1^{0.5}}}{{{n_1}}}} \right)^{\beta  - 1}},\]
${\kappa _0} = {m_p}{c^3}{B_ \odot }{\left( {{p_0}/({m_p}c)} \right)^{\beta  
- 1}}/\left( {32{\pi ^2}e{\nu _0}{E_{\nu 0}}} \right)$, $E_w(\nu \geq 
\nu_0)=4.35\times 10^{-3}$ erg cm$^{-3}$, and ${n_1} = n/{n_ \odot }$.

In our calculation, we take into account the particle diffusion along the 
magnetic 
field only, and, as mentioned above, we assume that 
$\kappa\equiv\kappa_\parallel$ . Particle diffusion across the magnetic field 
is not considered because  according to the results of the two-component 
(2D-slab) 
turbulence theory in the inner heliosphere, $\kappa_\perp\ll\kappa_\parallel$  
\citep{zank04}.

\begin{figure}
\plotone{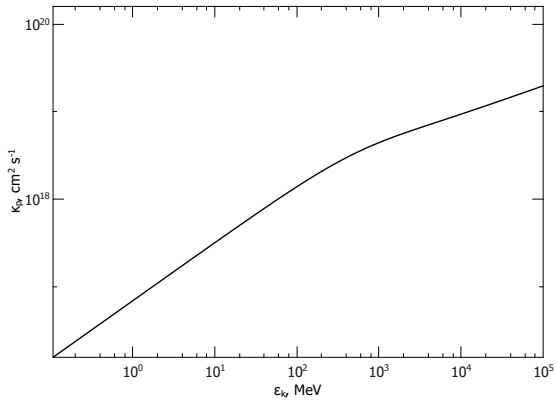} 
\figcaption{Dependence of the diffusion coefficient on energy for $\beta  
= 5/3$. 
\label{f3}}
\end{figure}

\begin{figure}
\plotone{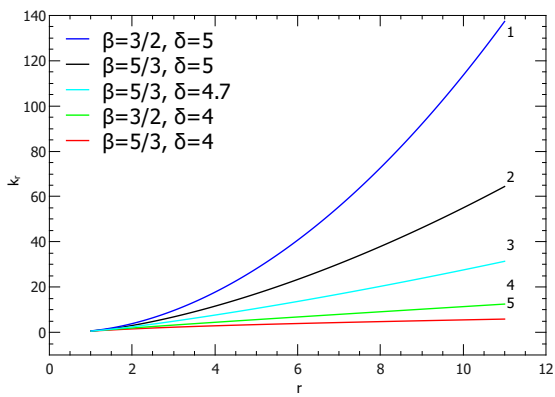} 
\figcaption{Dependence of the diffusion coefficient on radius for various 
$\beta $ and $\delta $. Curve 1 is for $\beta=3/2$ and $\delta  = 5$, curve 2 
is for $\beta = 5/3$ and $\delta = 5$, curve 3 is for $\beta = 5/3$ and 
$\delta  = 4.7$, curve 4 is for $\beta  = 3/2$ and $\delta  = 4$, and curve 5 
is for $\beta  = 5/3$, $\delta  = 4$.
\label{f4}}
\end{figure}

SEP spectra generation in the solar corona mainly depends on the diffusion 
coefficient. The properties of the Alfven turbulence spectrum used are, in 
fact, free parameters. They can be determined from the comparison of the model 
calculations to the observation results of SEPs. The dependences of the 
diffusion coefficient on energy for $\beta  = 5/3$  and on radius for 
various $\beta $ and $\delta $ are shown in Figures \ref{f3} and \ref{f4}, 
respectively.

\section{Results and discussion}

\begin{figure}
\plotone{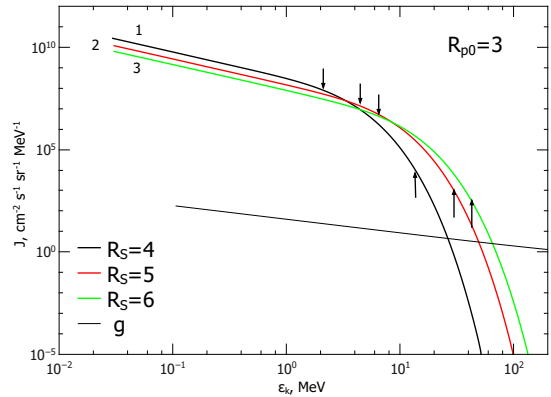} 
\figcaption{Particle intensity at the shock front as a function of kinetic 
energy for three shock radii of ${R_s} =$ 4, 5, 6, they are denoted by digits 1, 
2, and 3, respectively. The initial CME radius is $3{R_ \odot }$. The  
$\downarrow$ synbols mark the values of ${\varepsilon_m}(t)$ limiting the 
power-law spectrum region. The $\uparrow $ symbols mark the values of 
${\varepsilon_2}(t)$ denoting the cutoff region width. The thin 
line is the modulation parameter $g$. The parameter value scale is the same 
as on left axis of the figure.
\label{f5}}
\end{figure}

The formulated problem, Equations (\ref{eq1})--(\ref{eq6}) with edge 
conditions of $f\left( 
{r \to \infty ,p,t} \right) = 0$, and ${\left( {\partial f/\partial r} 
\right)_{|{R_p}}} = 0$ has been solved numerically. The solution algorithm
for the particle transport equation is similar to the one developed by 
\citet{byk96} to describe cosmic-ray acceleration in supernova remnants. For the
Alfven wave distribution with direction, we adopt $E_\nu ^ + /{E_\nu } = 0.7$, 
$E_\nu ^ - 
/{E_\nu } = 0.3$ and thus ${\delta _\nu } = 0.4$ in all calculation cases. 
Figure \ref{f5} shows the accelerated 
particle intensity of $J\left( {{\varepsilon _k}} \right) = {p^2}{f_p}$ at the 
shock front, which depends on the kinetic energy, for three shock front radii 
of ${R_s} = 
4,5,6$ with the following parameters: $\beta  = 5/3$, $\delta  = 5$, 
${\kappa _1} = {\kappa _p}{\kappa _r}$, ${\kappa _2} = {\kappa _p}$, ${V_p} = 
1000$ km s$^{-1}$, and ${R_{p0}} = 3$. We assume that the above parameters are 
standard 
and will only mention if they differ in the sequel. In the energy range of 
${\varepsilon _{\rm inj}} \leq {\varepsilon _k} \leq {\varepsilon _m}\left( t 
\right)$, a region of power-law spectrum is formed, the index
of which corresponds to the stationary spectrum of ${q_J} = \left( {{q_f} - 2} 
\right)/2 \approx 1.25$, where ${q_f} = 3{\sigma _{\rm ef}}/\left( {{\sigma 
_{\rm ef}} - 1} \right) = 4.5$ is the power-law index of the particle spectrum 
accelerated by the plane shock front with ${\sigma _{\rm ef}} = u'_1/u'_2 = 3$. 
The characteristic value of ${\varepsilon _m}( t )$ limiting the power-law 
spectrum region is equal to the energy of particles whose time of cyclic 
movement in the vicinity of the shock front equals to the average time $t_k$ 
defined in the 
Appendix, Equation~\ref{eqA4}. The value of ${\varepsilon _m}\left( t \right)$, 
indicated in Figure \ref{f5} by downward arrows, changes in time, as in the
acceleration by the plane shock front. The value of 
${\varepsilon_m}\left( t \right) = {\varepsilon _{\rm inj}}{\left( 
{{p_m}/{p_{\rm inj}}} \right)^2}$ is defined by Equation (\ref{eqA9}), in the 
Appendix. 

The decrease of the spectrum amplitude at the power-law spectrum region is 
caused by the spatial 
distribution of matter density in the solar atmosphere. The slight increase 
of injection energy is due to the increase of the shock speed (the lowest curve 
in 
Figure \ref{f1}). The cutoff region of the spectrum adjoins to the power region. 
The physical reason for the cutoff formation is the dispersion of the cyclic 
movement time relative to the average time $t_k$. The spectrum shape in the 
cutoff region is an important parameter since high-energy SEPs form this 
region. 
The values of ${\varepsilon _2}\left( t \right)$ determine the 
width of the cutoff region and are marked by the upward arrows in Figure 
\ref{f5}. Values of ${\varepsilon _2}$ are calculated according to equation 
(\ref{eqA7}). As can be seen from the figure, the ratio ${\varepsilon 
_2}/{\varepsilon _m}$ does not depend on time. Note that the ratio of 
$J\left( {{\varepsilon _2},t} \right)/J\left( {{\varepsilon _m},t} \right)$ 
calculated for the plane shock front exceeds the 
value from our calculation is approximately 10 times. The difference is 
apparently due to different
geometry. We calculate the particle acceleration up to ${R_s} = 6$. Then the
acceleration becomes ineffective due to a geometrical factor. The influence of 
the shock front size on particle acceleration is taken into account through the 
modulation parameter $g = R_s V_s/\kappa_1(R_s,\varepsilon_k )$. 
Particles with $g \leq 1$ intensively leave the vicinity of the shock front. It 
is the phenomenon of particle escaping which describes the injection of 
accelerated particles into the environment with monotonically changing 
parameters 
\citep{byk96}. The dependence of acceleration efficiency on the modulation 
parameter can be explained as follows: (1) the particles during cyclic 
movement move away from the shock front move away on a
distance of the diffusive 
length $L \approx \kappa _1 ( R_s,\varepsilon _k)/V_s$, and (2) the 
acceleration efficiency depends on the ratio between $L$ and ${R_s}$: if $L \ll 
{R_s}$, 
particles return to the shock front and are accelerated; if $L \geq R_s$, 
particles may not return and their acceleration is suppressed. In Figure 
\ref{f5} 
the thin curve shows the modulation parameter for ${R_s} = 6$. The 
parameter value scale is the same as on left axis of the figure.

\begin{figure}
\plotone{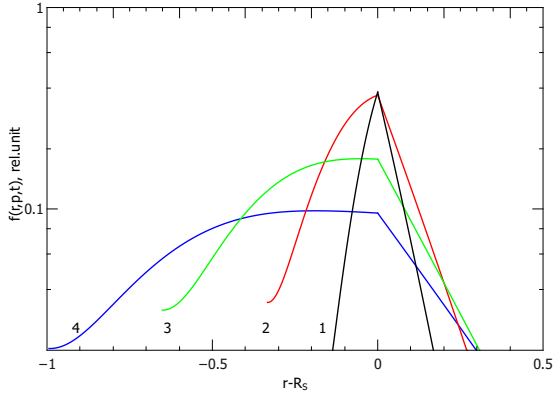} 
\figcaption{Spatial distribution of particle intensity in relative 
units. The digits 1, 2, 3, and 4 denote the particle intensity with 
${\varepsilon _k} = 1$ MeV for four shock radii ${R_s} =$ 3.5, 4, 5, and 6, 
respectively. The left boundary of the spatial distribution behind the shock 
front coincides with the piston radius.
\label{f6}}
\end{figure}

Figure \ref{f6} presents the spatial distribution of particle intensity in 
relative units with ${\varepsilon_k} = 1$ MeV for $R_s = $ 3.5, 4, 5, and 6. As 
we 
can see from Figure \ref{f5} the particle intensity for this energy at ${R_s} = 
4$ has almost reached the stationary value. Accordingly, its spatial 
distribution is similar to that of the stationary shape: the distribution is 
exponential 
ahead of the shock front, and an interval of constant value forms behind 
the shock front. In the subsequent expansion, the shape of the spatial 
distribution ahead of 
the shock front remains the same. The expansion of the volume filled with 
particles 
is caused by the increase of the diffusion coefficient at the shock front. The 
region of constant value behind the front increases with time. The left 
boundary of the intensity distribution behind the shock front coincides with 
the 
piston radius. The monotonic decrease of the intensity at the shock front at 
${R_s} 
> 4$ is caused by the decrease of the injection rate.

\begin{figure}
\plotone{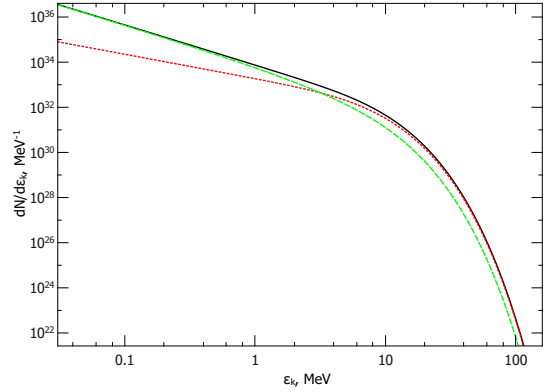} 
\figcaption{Spectrum of the total accelerated particle number as a function 
of kinetic energy at ${R_s} = 6$. The dashed curve represents the spectrum 
behind the shock front, the dotted curve represents the spectrum ahead of the 
shock front, and the solid curve is their sum of them and represents the 
spectrum of the total particle number.
\label{f7}}
\end{figure}

In Figure \ref{f7}, the differential spectrum of the total number of 
accelerated particles is plotted as a function of kinetic energy, which is 
defined as
$dN/d{\varepsilon _k} = (4\pi p^2 /V) d\Omega \mathop \smallint_V 
f_p (p,r,t) r^2 dV$, where $V$ is the volume at ${R_s}= 6$ per the unit of 
a solid angle $(d\Omega = 1)$. The curves in the figure correspond to the 
spectra of the
total accelerated particle number ahead of and behind the shock front as well 
as 
their sum. The distribution depends on particle energy: there 
are more particles with energies of ${\varepsilon _{\rm inj}} < {\varepsilon _k} 
< 3$ MeV behind the shock front, and the opposite for particles 
with 
energies of ${\varepsilon _k} > 3$ MeV.

\begin{figure}
\plotone{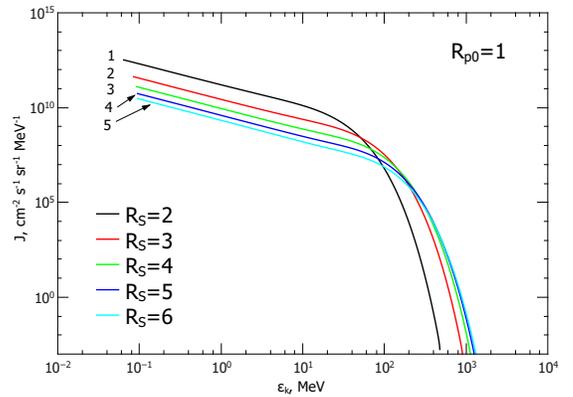} 
\figcaption{Particle intensity at the shock front depending on kinetic energy 
for five shock radii $R_s=$ 2, 3, 4, 5, 6 denoted by digits 1--5, 
respectively. The initial CME radius is $1 R_\odot$.
\label{f8}}
\end{figure}

\begin{figure}
\plotone{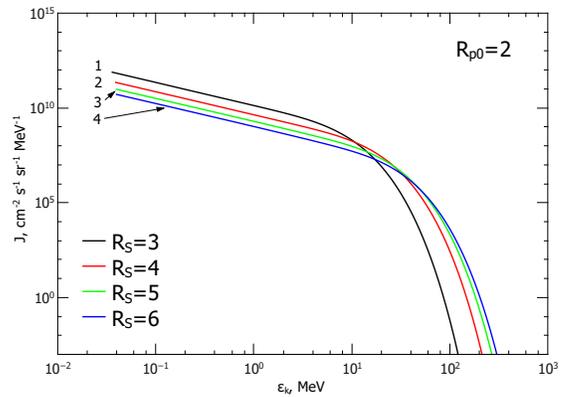} 
\figcaption{Particle intensity at the shock front depending on kinetic energy 
for four shock radii $R_s =$ 3, 4, 5, 6 denoted by digits 1--4, 
respectively. The initial CME radius is $2 R_\odot$. 
\label{f9}}
\end{figure}

Figures \ref{f8}, and \ref{f9} show the particle intensities at the shock front 
as a
function of energy for initial radii of $R_{p0}=$ 1, and 2, respectively. One 
can see that the smaller $R_{p0}$ is, the higher is $V_s$ (see 
Figure \ref{f1}), and the more efficient is the particle acceleration. From 
Figures 
\ref{f2}, \ref{f5}, \ref{f8}, and \ref{f9}, one can conclude that the
spectrum is formed 0.2--1 hr after the beginning of CME motion. At that time 
${R_p} =$ 3~--~5 and ${R_s} =$ 5~--~10, and may begin an injection of particles 
into interplanetary medium, which is in agreement with observations.

\begin{figure}
\plotone{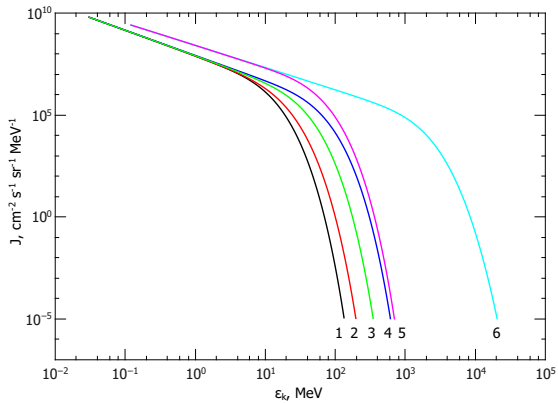} 
\figcaption{Particle intensity at the shock front depending on kinetic energy 
for ${R_s} = 6$ and different parameters. See the text for details.
\label{f10}}
\end{figure}

Figure \ref{f10} shows the intensity of particles at the shock front for 
different values of parameters. All calculations here start at $R_{p0}=3$ 
and finish when ${R_s} = 6$. The intensities marked by digit 1 in 
Figure \ref{f10} and digit 3 in Figure \ref{f5} are calculated with 
standard parameters. Curves 2--5 differ in one parameter: 2 is for 
$\kappa_2=0.1 \kappa_p$, 3 is with $E_w(\nu\geq\nu_0) = 8.7\times 10^{-3}$ erg 
cm$^{-3}$, 4 is for $\delta  = 4$; and 5 is with ${V_p} = 2000$ km s$^{-1}$. 
Curve 6 represents the total influence of the changes.
The particle acceleration rate in the regular acceleration depends on the 
value of $\kappa /V_s^2$ and it can explain the intensity changes in Figure 
\ref{f10}. Here, the parameter $\kappa$ is proportional to the sum of 
diffusion coefficients ahead of and behind the shock front.

In most cases SEP flows are measured at the Earth's orbit. Therefore, it is 
necessary to somehow connect spectra of particles accelerated in the solar 
atmosphere and ones registered in interplanetary space. The CME itself also 
influences particle propagation in interplanetary space. The extent of 
the influence is 
determined by the modulation parameter $g = R_s V_s/\kappa_{\rm ip} 
(R_s,\varepsilon_k)$, where $\kappa_{\rm ip}$ is the spatial diffusion 
coefficient in interplanetary space. Depending on the value of $g$ ((1) $g 
\gg 
1$, (2) $g\sim 1$, and (3) $g\ll 1$) there are three possible scenarios for 
particle propagation. In the first case, the CME influences the form of the 
spectrum and particle spatial distribution. In observations, the first case 
describes ESPs (particles with $\varepsilon_k \leq 10$ MeV the 
flow of which after a plateau reaches a maximum at the shock front). In the 
second case,
the CME only influences the particle spatial distribution. The second scenario 
describes particles with $10 <\varepsilon_k < 50$ MeV nucleon$^{-1}$ in 
observations 
and their constant flow value retains to CME arrival \citep{rea90}. In the 
third case, the CME does not affect particle propagation. In observations, 
these 
SEPs are particles with ${\varepsilon_k} \geq 100$ MeV nucleon$^{-1}$.
A ``black box" model \citep{kall97} is widely used in order to determine SEP 
flows in gradual events. The particle distribution in interplanetary space from 
the source can be obtained from the model; the source is assigned a particle 
spectrum at the moving shock front \citep{hera92, hera95, kall97, lario98, 
kall99}. The model does not take into account the correlation of particle flow 
characteristics with the phenomena occurring at the shock front. Detailed 
dynamic 
and self-consistent models of the shock propagation and particle acceleration 
have been developed by \citet{zank00} for strong shock waves 
and by \citet{rice03} for shock waves with arbitrary intensities. 
Using these models, \citet{li03} have invented a numerical method known 
as Particle Acceleration and Transport in the Heliosphere (PATH) to simulate 
SEP events in interplanetary space. The model includes local particle injection 
at the moving quasi-parallel shock wave, diffusive shock acceleration, 
self-consistent Alfven wave generation by accelerated particles, particle 
trapping and escape from the complex shock, and particle propagation in the 
inner 
heliosphere. Using the PATH method, the characteristics of heavy nucleus flow 
\citep{li05, zank07}, and the dependence of particle flow 
characteristics on the angle between the magnetic field and the normal to the 
shock front, including heavy nucleus \citep{li09, li12}, have been 
calculated. The 
diffusion coefficient used was determined from the two-component (2D-slab) 
turbulence theory \citep{zank04, zank06}. The distribution depends on particle 
energy. The comparison of the calculated 
results with specific events \citep{verkh09, verkh12}, including a mixed 
population of both flare and shock-accelerated particles \citep{verkh10}, shows 
their general agreement. The agreement demonstrates that the PATH model takes 
into 
account the main physical factors determining SEP acceleration and propagation.
In this paper, we 
will consider the third scenario only. 
The simplified approach of particle propagation is formulated following 
\citet{bt03}. Particle propagation is described by the transport equation for 
the distribution function of $f(r,p,t)$ in diffusive approximation:
\[\frac{{\partial f}}{{\partial t}} = \frac{1}{{{r^2}}}\frac{\partial 
}{{\partial r}}\left( {\kappa {r^2}\frac{{\partial f}}{{\partial r}}} \right) + 
Q,\] 
where $Q=F(p)/(\Omega_s R_f^2) \delta(t-t_f)$ is the source term and 
$F(p)/\Omega_s$ is the spectrum of the total particle number accelerated in the 
solar atmosphere per unit solid angle. The source term shows that the 
particles accelerated by the time ${t_f}$ occupy a volume with radius 
${R_f}$ and will instantly be injected into the surrounding medium when $t = 
{t_f}$. The 
solution of the equation at $r \gg {R_f}$ and $t > {t_f}$ is as follows 
\citep{kri65}: \[f\left( {r,p,t} \right) = \frac{{F\left( p 
\right)}}{{2{{{\Omega }}_s}r_e^3}}\frac{1}{{t_*^3}}{e^{ - {r_*}/{t_*}}},\] 
where 
$t_*=t/T$, ${r_*} = r/{r_e}$, and $T = r_e^3/\kappa_{\rm ip,e}$. In the 
calculations, the 
expression $\kappa \propto r$ \citep{bt03} is used. 
The maximum of the distribution function at ${r_*} = 1$ occurs at $t_{*, \rm 
max} = 
1/3$. As a result, we can derive the spectrum of maximal intensities as a
function of kinetic energy:
\begin{equation}
J_{\rm max} = {p^2}f\left( {{r_*} = 1,p,{t_{*,max}}} \right) = 
0.67\frac{{{p^2}F\left( p \right)}}{{{{\Omega }_s}r_e^3}}. \label{eq7}
\end{equation}

\begin{figure}
\plotone{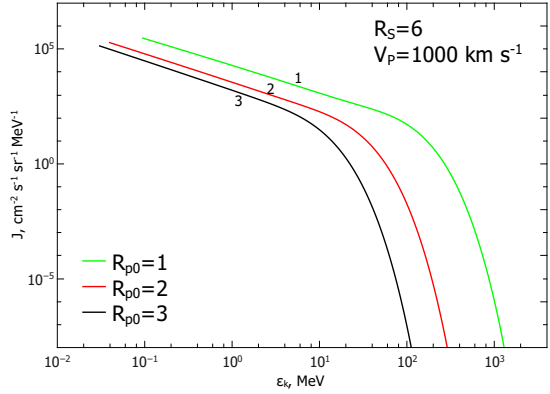} 
\figcaption{Intensity of the maximum values of particles' flux depending on 
kinetic energy at the Earth's orbit. Curves marked by digits 1, 2, and 3 
correspond to CME radii $R_{p0}=$ 1, 2, and 3, respectively.
\label{f11}}
\end{figure}

Figure \ref{f11} shows the intensity of the maximum values in depending on the 
kinetic 
energy at the Earth's orbit, according to Equation (\ref{eq7}). For the 
injected 
particles, we use the spectra of the total number of particles at ${R_s} 
= 6$ for the three variants shown in Figures \ref{f5}, \ref{f8}, and \ref{f9}. 
One can see from Figure \ref{f11} that the smaller $R_{p0}$ is, and accordingly 
the higher  
$V_s$ is, the higher are the particles' flux and maximum energies.

\begin{figure}
\plotone{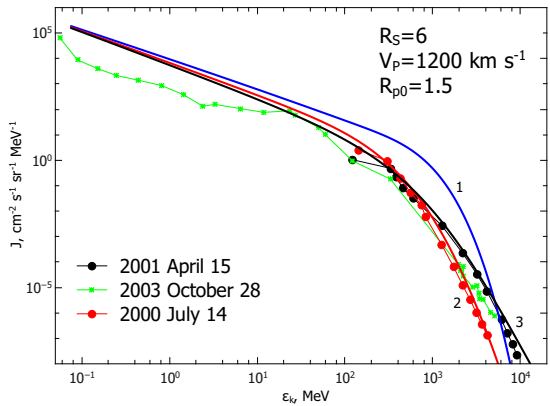} 
\figcaption{Intensity of the maximal values of the SEP flux as a function of 
kinetic 
energy at the Earth's orbit. The following symbols denote measurements: 
asterisks 
are from 2003 October 28 \citep{kry15}, red circles are from 2000 July 14, and
black circles 
are from 2001 April 15 \citep{bom07}. 
Calculation curves correspond to the following parameters: (1) $\beta  = 
5/3$, $E_w (\nu \geq \nu _0) = 8.7\times 10^{-3} $ erg cm$^{-3}$; (2) 
$\beta  
= 2.2$, $E_w (\nu \geq \nu _0) = 3.05\times 10^{-2} $ erg cm$^{-3}$; and (3) 
$\beta  = 2.4$, $E_w (\nu \geq \nu _0) = 1.1\times 10^{-1}$ erg cm$^{-3}$.
\label{f12}}
\end{figure}

Figure \ref{f12} presents the SEP intensity depending on the energy at the 
Earth's orbit 
for three GLE events \citep{bom07, kry15}. As we know from observations, ${V_p} 
= 1200$ km s$^{-1}$ and ${R_{p0}} = 1.5$ for the event on 2001 April 
15 \citep{bom07}. We use the following values in the calculation: $\delta  = 
4$, ${\kappa _1} = 
{\kappa _p}{\kappa _r}$, ${\kappa _2} = {\kappa _p}$, and ${V_p} = 1200$ km 
s$^{-1}$. The process starts from $R_{p0} = 1.5$ and terminates when
${R_s} = 6$. We use $\beta  = 5/3$ and ${E_w}\left( {\nu  \geq
{\nu _0}} \right) = 8.7\times 10^{-3}$ erg cm$^{-3}$ to calculate curve 1, 
shown 
in Figure \ref{f12}. Apparently, the maximal energy in the spectrum agrees with 
the observations; however, the cutoff shape significantly differs from the 
registered one. 

The shape of the spectrum near cutoff energies is determined by the dependence 
of the
diffusion coefficient on energy. According to equation (\ref{eq6}), $\kappa_p 
\propto [p/(m_p c)]^{2-\beta}$ for relativistic energies; therefore, the 
diffusion coefficient decreases with the increase of particle energy if $\beta  
> 
2$. However, the increase of the wave spectrum power index causes the increase 
of the
diffusion coefficient at low energies and suppresses acceleration efficiency. 
Thus, the increase of $\beta $ requires the increase of $E_w(\nu \geq \nu_0)$. 
Curves 2 and 3 in Figure \ref{f12} are calculated with $\beta  = 2.2$, $E_w (\nu 
\geq \nu _0) = 3.05\times 10^{-2} $ erg cm$^{-3}$ and $\beta  = 2.4$, $E_w (\nu 
\geq \nu _0) = 1.1\times 10^{-1}$ erg cm$^{-3}$, respectively. One can see that
in these cases, the calculated spectra agree with the registered ones. However, 
the assumed values of $E_w(\nu \geq \nu _0)$ significantly exceed the standard 
values. We can suggest some possible reasons for the agreement with the 
standard values: 
(1) the values of $\nu_1$, and $\nu_0$ in the solar atmosphere are an order of 
magnitude greater 
than the used ones; (2) the wave spectrum 
in the inertial region has a more complicated dependence on frequency---for 
example, it consists of two parts with different power-law indexes; and (3) the 
wave 
energy increases due to the generation by accelerated particles. We do not 
discuss 
here the difference between the calculated and measured spectra at low energies 
$\varepsilon _k < 100$ MeV because the Krimigis model does not describe the
interplanetary propagation of particles with such energies. 

\section{Conclusion}

The relationship between the CME and the shock speeds moving in the solar 
atmosphere is defined from the solution of the gas-dynamic equations. The shock 
speed increases with the decrease of the initial CME radius. The accelerated 
particle spectra as a function of time has been reproduced by numerical 
solution of the diffusive transport equation with a set of realistic parameters. 
Depending on the initial CME radius, its speed, and Alfven 
wave magnetic energy for $\beta  = 5/3$, the accelerated particle spectrum is 
established at 10~--~60 minutes after the beginning of CME motion. The maximum
energies of the particles reach 0.1--10 GeV. By that time, the CME radii are 
3~--~5 $R_\odot$ and the shock front radii are 5~--~10 ${R_ \odot }$, which
agree with observations. The calculation results and observations are in 
agreement 
if $\beta > 2$. However, in this case the Alfven wave magnetic energy is 
significantly higher than the standard one. 

\acknowledgements 
The authors thank E.G. Berezhko for suggesting the research topic. The 
research was 
supported by the Russian Science Foundation (Project No. 14-12-00760). 

\appendix
\section{Particle Acceleration by Plane Shock Front}

To estimate the accuracy of the numerical algorithm, we calculate the particle 
acceleration by the plane shock front, which has a constant speed ${V_s}$ and 
moves in infinite medium. The corresponding particle transport equation for an
isotropic distribution function $f_i(x,p,t)$ is
\[\frac{{\partial {f_i}}}{{\partial t}} = {\kappa _i}\frac{{{\partial 
^2}{f_i}}}{{\partial {x^2}}} - {w_i}\frac{{\partial {f_i}}}{{\partial x}} + 
\frac{{{N_0}{u_1}}}{{4\pi p_{\rm inj}^2}}\delta \left( {p - {p_{\rm inj}}} 
\right)\delta 
\left( {x - {x_s}} \right)H\left( t \right),\]
where the subscript $i$ can be 1 or 2, corresponding to the regions ahead of 
$(x > 
{x_s})$ and behind $(x < {x_s})$ the shock front, respectively; $\kappa_1$ and 
$\kappa_2$ are the spatial diffusion coefficients; $w_1$ and ${w_2} = \left( 
{\left( {\sigma  - 1} \right){V_s} + {w_1}} \right)/\sigma $ are the flow 
velocity ahead of and behind the shock front; $\sigma$ is the compression 
ratio; 
${x_s} = {x_{s0}} + {V_s}t$ is the front position; ${u_1} = {V_s} - {w_1}$; 
${N_0}$ is the particle number density injecting at the momentum $p_{\rm inj}$; 
and 
$H$ is the Heaviside function. In the case when the coefficients $\kappa_1$ and 
$\kappa_2$ are constants and relate to each other as $\kappa_1/\kappa_2 = 
\sigma^2$, the task has an exact solution:
\begin{equation}
{f_1}/{f_\infty } = 0.5 {\rm erfc}\left( {\sqrt {\frac{{{t_0}}}{t}} \left( 
{a1 + a3} 
\right)/4 - \sqrt {\frac{{{t_0}}}{t}} } \right) + 0.5{\left( 
{\frac{p}{{{p_{\rm inj}}}}} \right)^{a3}}{\rm erfc}\left( {\sqrt 
{\frac{{{t_0}}}{t}} 
\left( {a1 + a3} \right)/4 + \sqrt {\frac{{{t_0}}}{t}} } \right), \label{eqA1} 
\end{equation}

\[{f_2}/{f_\infty } = 0.5{\rm erfc}\left( {\sqrt {\frac{{{t_0}}}{t}} \left( 
{a2 + a3} 
\right)/4 - \sqrt {\frac{{{t_0}}}{t}} } \right) + 0.5{\left( 
{\frac{p}{{{p_{\rm inj}}}}} \right)^{a3}}{e^{a2}}{\rm erfc}\left( {\sqrt 
{\frac{{{t_0}}}{t}} \left( {a2 + a3} \right)/4 + \sqrt {\frac{{{t_0}}}{t}} } 
\right),\] 
where ${f_i}\left( {x,p,t} \right)/{f_\infty }$ is the relative spectrum; 
${f_\infty } = f\left( {{p_{\rm inj}}} \right){\left( {p/{p_{\rm inj}}} 
\right)^{ - 
q}}$, with $q = 3\sigma /\left( {\sigma  - 1} \right)$, is the stationary 
spectrum of accelerated particles at the shock front; ${t_0} = 4{\kappa 
_1}/u_1^2$; ${u_2} = {u_1}/\sigma $; ${a_1} = {u_1}\left( {x - {x_s}} 
\right)/{\kappa _1}$; ${a_2} = {u_2}\left( {{x_s} - x} \right)/{\kappa _2}$; 
${a_3} = 3\left( {\sigma  + 1} 
\right)/\left( {\sigma  - 1} \right)\ln \left( {p/{p_{\rm inj}}} \right)$; and
${\rm erfc}$ is 
the additional probability integral. To derive the above expressions, the 
Laplace transformation in time is used.

\begin{figure}
\plotone{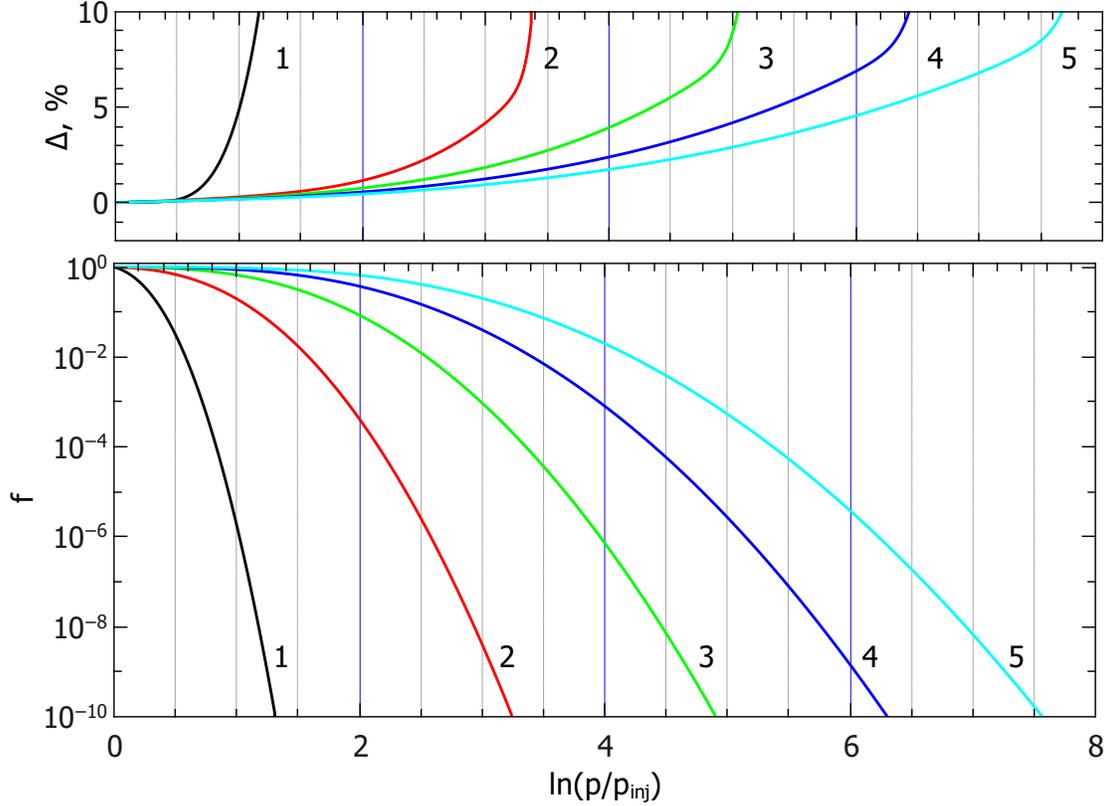} 
\figcaption{Particle spectrum at the shock front as a function of momentum 
for five successive time moments. $f=f_1(x_s, p_*, t)/f_\infty$ is the relative 
particle spectrum and ${f_\infty }$ is the stationary spectrum. On the top 
panel, 
$\Delta = \left( {f - {f_{\rm ex}}} \right)/{f_{\rm ex}}\times100\% $ is the 
deviation 
of the relative spectrum of the numerical solution from the exact solution in 
percent; 
${f_{ex}}$ is the relative spectrum of the exact solution.
\label{f13}}
\end{figure}

\begin{figure}
\plotone{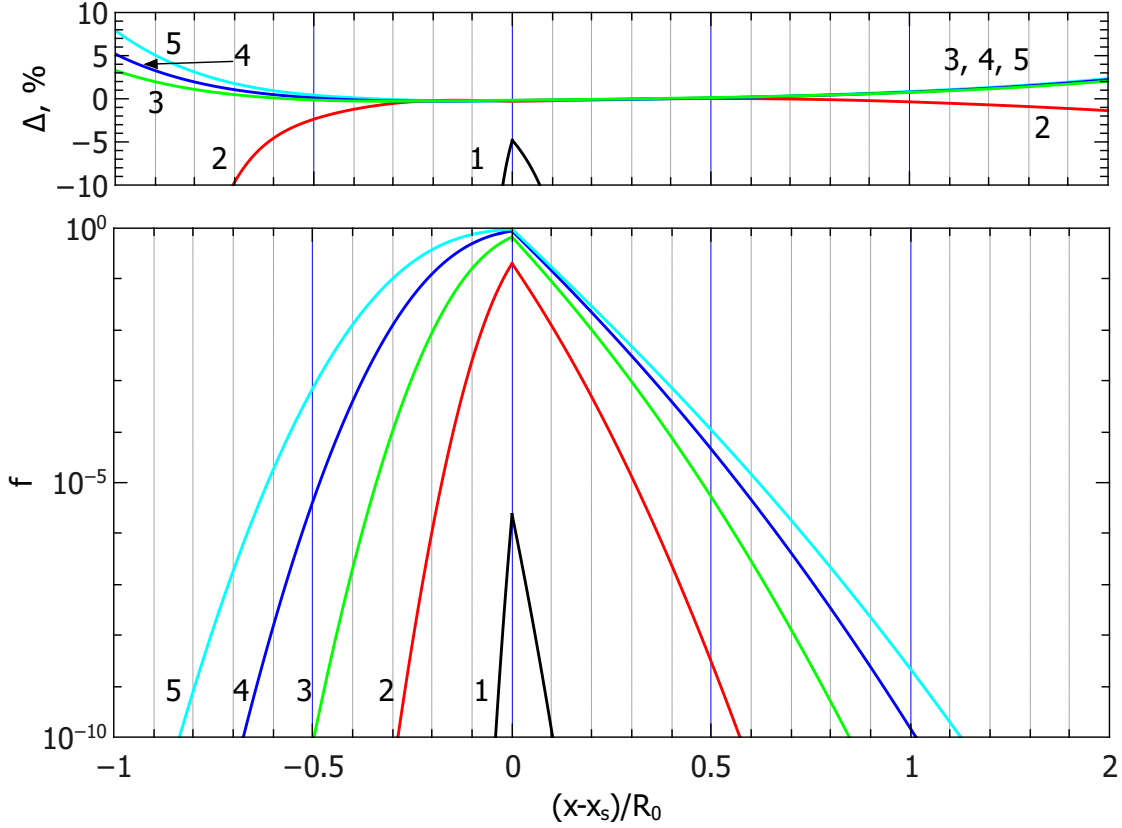} 
\figcaption{Spatial distribution of the particle spectrum for five successive 
time moments. $f=f_1(x>x_s, p_*, t)/f_\infty$, $f=f_2(x<x_s, p_*, t)/f_\infty$, 
and
$\ln(p_*/p_{\rm inj}=1$. On the top panel, $\Delta  = \left( {f - {f_{\rm ex}}} 
\right)/{f_{\rm ex}}\times 100\% $ is the deviation of the spatial spectrum 
dependence 
of the numerical solution from the exact solution in percent; ${f\rm ex}$ is 
the spatial 
distribution of the relative spectrum of the exact solution.
\label{f14}}
\end{figure}

The solution (\ref{eqA1}) at the shock front $x = x_s$ is similar to that
obtained in \citet{ber88} and \citet{axf81}. Figure \ref{f13} presents the 
relative 
spectrum of accelerated particles at the shock front $ {f = {f_1}\left( 
{{x_s},p,t} \right)/{f_\infty }}$ as a function of momentum for five 
successive time instants. Figure \ref{f14} shows the spatial 
distribution of the relative particle spectrum ($f=f_1(x>x_s, p_*, 
t)/f_\infty$, $f=f_2(x<x_s, p_*, t)/f_\infty$) whose momentum logarithm is 1 
($\ln \left( {{p_*}/{p_{\rm inj}}} \right) = 1$) as a function of distance for 
five successive time moments. The deviations of the corresponding values of the 
numerical 
solution from the exact solution in percent are given at the top panels of the
figures.

One can see from Figures \ref{f13} and \ref{f14} that the accuracy of the 
calculation 
depends on the amplitude of the relative spectrum. The relative deviation does 
not 
exceed a few percent if the value is higher than $10^{-6}$. In general, the 
comparison confirms the sufficient accuracy of the numerical calculation 
algorithm. 

The exact solution (\ref{eqA1}) describes the formation of the accelerated 
particle 
spectrum in time. The shape of the spectrum depends on the momentum. The shape 
is 
mainly determined by the first term in equation (\ref{eqA1}). The value of 
${p_m}\left( t \right)$, which separates different shapes of the spectrum, can 
be 
found from (\ref{eqA1}) by equating the argument of the first additional 
probability integral to zero. Hence, following this equation, 
\[f\left( {{x_s},{p_m},t} \right)/{f_\infty }\left( {{p_m}} \right) \approx 0.5\]
and
\begin{equation}
\frac{p_m(t)}{p_{\rm inj}} = {\text{exp}}\left[ {\frac{{\left( {\sigma  
- 
1} \right)}}{{3\left( {\sigma  + 1} \right)}}\frac{{u_1^2t}}{\kappa_1}} 
\right].  
\label{eqA2}
\end{equation}
The value of ${p_m}\left( t \right)$ separates the momentum region of ${p_{inj}} 
\leq p \leq {p_m}\left( t \right)$, where the spectrum is close to a
stationary one, and the region $p > {p_m}\left( t \right)$, where the greater 
the momentum, the more the spectrum deviates from the stationary one. The 
width of the cutoff region can be determined by equating the argument of the 
first term in Equation \ref{eqA1} to the value of A;
\[\sqrt {\frac{{{t_0}}}{t}} \frac{3}{4}\frac{{\left( {\sigma  + 1} 
\right)}}{{\left( {\sigma  - 1} \right)}}\ln \frac{{{p_2}}}{{{p_{\rm inj}}}} - 
\sqrt 
{\frac{{{t_0}}}{t}}  = A,\]
and therefore
\begin{equation}
\frac{{{p_2}}}{{{p_m}}} = {\text{exp}}\left[ {A\sqrt {\frac{{4\left( {\sigma  + 
1} \right)}}{{3\left( {\sigma  - 1} \right)}}\ln \frac{{{p_m}}}{{{p_{\rm 
inj}}}}} } 
\right], \label{eqA3}
\end{equation}
where Equation \ref{eqA2} is used. Here $A$ defines the deviation value of the 
spectrum 
from the stationary one $f\left( {{p_2},t} \right)/{f_\infty }\left( {{p_2}} 
\right) \approx {\rm erfc}\left( A \right)$ at momentum ${p_2}$. 
The width of the 
cutoff region increases with time. The temporary dynamics of the spatial 
distribution, shown in Figure \ref{f14}, is the same for all momenta, differing 
by the offset. Figures \ref{f13} and \ref{f14} show that the spatial 
distribution 
in the region ahead of the shock front becomes exponential, with an interval of 
constant value being formed in the region behind the shock front. The 
subsequent changes in the spatial distribution are only an expansion of the 
interval.

Let us consider the approximate solution of the problem of particle acceleration 
by the plane shock front when the diffusion coefficients depend on momentum. 
Such a solution can be used to interpret the particle spectrum dependence 
on task parameters.

It is known that the acceleration of particles by the regular mechanism is 
caused by the cyclic particle movement in the vicinity of the shock 
front \citep{kry77, bell78}. Statistical characteristics of the 
movement, such as the average time of ${t_k}$ that particles spend on $k$ 
cycles, and the corresponding dispersion is $d_k^2$, are as follows 
\citep{for83,ber88}:
\begin{equation}
t_k = \frac{3}{{{u_1} - {u_2}}}\int_{p_{\rm inj}}^{p_k} \left( {{\kappa_1}/{u_1} 
+ {\kappa_2}/{u_2}} 
\right)\frac{{dp}}{p}, \,\,\,\, d_k^2 = 6/\left( {{u_1} - {u_2}} \right) 
\int_{p_{\rm inj}}^{p_k} \left( 
{\kappa_1^2/u_1^3 + \kappa_2^2/u_2^3} \right)\frac{{dp}}{p}. \label{eqA4}
\end{equation}
From the comparison of ${t_k}$ from equation (\ref{eqA4}) for 
${\kappa_1}/{\kappa_2} = 
{\sigma ^2}$ with its counterpart from Equation (\ref{eqA2}), we get ${t_k} = 
t$ 
and 
${p_k} = {p_m}$. Hence, the characteristic value of ${p_m}\left( t \right)$ is 
equal to the particle momentum, the time of a cyclic movement which 
is equal to the average time $t_k$.

Using the central limit theorem of probability theory and ${t_k}$, $d_k^2$, one 
can find the particles' distribution function \citep{ber88}:
\begin{equation}
	\frac{f}{{{f}_{\infty }}}=0.5~{\rm erfc}\left( \sqrt{\frac{{{\delta 
}_{1}}{{t}_{k}}}{t}}- \sqrt{\frac{{{\delta }_{1}}t}{{{t}_{k}}}} 
\right)+0.5{{e}^{4{{\delta }_{1}}}}~{\rm erfc}\left( \sqrt{\frac{{{\delta 
}_{1}}{{t}_{k}}}{t}}-\sqrt{\frac{{{\delta }_{1}}t}{{{t}_{k}}}} \right),  
\label{eqA5}
\end{equation}
where ${{\delta }_{1}}=t_{k}^{2}/2d_{k}^{2}$. If in the approximate solution 
$\kappa_1/\kappa_2=\sigma^2$, then the solution from Equation (\ref{eqA5}) 
according to Equation (\ref{eqA4}) is the same as that for Equation 
(\ref{eqA1}). If 
${{\kappa}_{1}}={{\kappa}_{10}}{{\left( 
p/{{p}_{\rm inj}} \right)}^{\alpha }}$, ${{\kappa}_{2}}={{\kappa}_{20}}{{\left( 
p/{{p}_{inj}} 
\right)}^{\alpha }}$, and ${{\left( p/{{p}_{inj}} 
\right)}^{\alpha }}\gg 1$, it follows from Equation (\ref{eqA4}) that
\begin{equation}
	{{t}_{k}}=\frac{{{\kappa}_{10}}}{u_{1}^{2}}\frac{q}{\alpha }\left( 
1+\sigma 
\frac{{{\kappa}_{20}}}{{{\kappa}_{10}}} \right){{\left( 
\frac{{{p}_{k}}}{{{p}_{\rm inj}}} 
\right)}^{\alpha }}, \,\,\,\, d_{k}^{2}={{\left( 
\frac{{{\kappa}_{10}}}{u_{1}^{2}} 
\right)}^{2}}\frac{q}{\alpha }\left( 1+{{\sigma }^{3}}{{\left( 
\frac{{{\kappa}_{20}}}{{{\kappa}_{10}}} \right)}^{2}} \right){{\left( 
\frac{{{p}_{k}}}{{{p}_{\rm inj}}} \right)}^{2\alpha }}.  		
\label{eqA6} 
\end{equation}

The solution from equation (\ref{eqA5}) is similar to the one from equation 
(\ref{eqA1}). The values of ${p_m}$ and ${p_2}$ from equation (\ref{eqA5}) are 
obtained the same way as in Equation (\ref{eqA1}). The result is
\begin{equation}
\frac{{{p_m}}}{{{p_{\rm inj}}}} = {\left( {\alpha u_1^2t/\left( 
{q{\kappa_{10}}\left( 
{1 
+ \sigma \frac{{{\kappa_{20}}}}{{{\kappa_{10}}}}} \right)} \right)} 
\right)^{1/\alpha }}, 
\,\,\,\, \frac{{{p_2}}}{{{p_m}}} = y_*^{2/\alpha },\,\,\,\, \frac{{{\varepsilon 
_2}}}{{{\varepsilon _m}}} = {\left( {\frac{{{p_2}}}{{{p_m}}}} \right)^2},
\label{eqA7}
\end{equation}
where ${{y}_{*}}=0.5B+\sqrt{1+{{\left( 0.5B \right)}^{2}}}$, 
$B={\left[ 2\alpha \left( 1+{{\sigma }^{3}}{{\left( 
{{\kappa}_{20}}/{{\kappa}_{10}} 
\right)}^{2}} \right)/(q{{\left( 1+\sigma {{\kappa}_{20}}/{{\kappa}_{10}} 
\right)}^{2}} 
\right]^{0.5}}A$.

In the case of spatial dependence of the diffusion coefficients on 
radius, we can generalize expression (\ref{eqA7}). Taking into account the 
definition of $p_m (t)$, it is possible to write the equation
\begin{equation}
\frac{d{{p}_{m}}}{dt}=\frac{\Delta p}{\langle t \rangle 
}={{p}_{m}}\frac{\left( {{u}_{1}}-{{u}_{2}} \right)}{3\left( 
{{\kappa}_{1}}/{{u}_{1}}+{{\kappa}_{2}}/{{u}_{2}} \right)},
\label{eqA8}
\end{equation}
where $\langle t \rangle  = 4({{\kappa_1}/{u_1} + {\kappa_2}/{u_2}})/v$ is the 
average time of one cycle of a particle; $\Delta 
{p_m} = 4{p_m}\left( {{u_1} - {u_2}} \right)/3v$ is the average momentum change 
of a particle over one cycle; and $v$ is the particle speed \citep{ber88}. 
For the exact solution $\left( {{\kappa_1}/{\kappa_2} = {\sigma ^2}} \right)$ 
with 
diffusion coefficients depending on the momentum of $(\kappa\propto p^\alpha)$, 
after 
integrating equation (\ref{eqA8}), one can obtain Eqs (\ref{eqA2}) and 
(\ref{eqA7}). We represent the parameters by ${\kappa_1} = 
{\kappa_{10}}{\left( {p/{p_{\rm inj}}} \right)^\alpha }{\left( {r/{R_0}} 
\right)^{{d_1}}}$, ${\kappa_2} = {\kappa_{20}}{\left( {p/{p_{\rm inj}}} 
\right)^\alpha 
}$, 
where ${\kappa_1}$ is the diffusion coefficient ahead of the shock front, 
${\kappa_2}$ is 
the counterpart behind the shock front; $R_0$ is the spatial scale; and ${R_s} 
= 
{R_{S0}}{\left( {t/{t_0}} \right)^{{d_2}}}$ is the shock radius depending on 
time. In this case the diffusion coefficient at the shock front is ${\kappa_1} 
= 
{\kappa_*}{\left( {p/{p_{inj}}} \right)^\alpha }{\left( {t/{t_0}} 
\right)^{{d_1}{d_2}}}$, where ${\kappa_*} = {\kappa _{10}}{\left( 
{{R_{S0}}/{R_0}} 
\right)^{{d_1}}}$. Substituting it into equation (\ref{eqA8}) and dividing the 
variables, one can obtain 
\begin{equation}
\frac{{{p_m}}}{{{p_{\rm inj}}}} = {\left[ {\frac{{\alpha 
u_1^2}}{{q{\kappa_*}}}\mathop 
\smallint \limits_{{t_0}}^t \frac{{dt}}{{{{\left( {\frac{t}{{{t_0}}}} 
\right)}^{{d_1}{d_2}}} + \alpha }}} \right]^{1/\alpha }},
\label{eqA9}
\end{equation}
where ${({p_m}/{p_{\rm inj}})^\alpha } \gg 1$, $\alpha  = \sigma {\kappa 
_{20}}/{\kappa _*}$. Equations (\ref{eqA3}) and (\ref{eqA5}) show that 
${p_2}/{p_m}$ depends on ${\delta_1} = t_k^2/2d_k^2$, which, according to 
equation (\ref{eqA4}), is determined by the ratio of the squares of the 
diffusion 
coefficients. Thus, one may assume that in this case the width of the cutoff 
region will still be defined by Equation (\ref{eqA7}).

\end{document}